\documentclass{cernyrep}
\usepackage{texnames}
\usepackage[T1]{fontenc}
\usepackage{placeins}
\usepackage{varwidth}
\usepackage{xcolor}
\usepackage{cite}

\begin{document}
\title{Protection Related to High-power Targets}
\author{M.A. Plum}
\institute{Oak Ridge National Laboratory, Oak Ridge, Tennessee, USA}

\maketitle % this produces the title block

% Define some shortcuts
 \newcommand{\Hm}{H$^-$ }
 \newcommand{\Hn}{H$^0$ }
 \newcommand{\Hp}{H$^+$ }

\begin{abstract}
Target protection is an important part of machine protection.
The beam power in high-intensity accelerators is high enough that a single wayward pulse can cause serious damage.
Today's high-power targets operate at the limit of available technology, and are designed for a very narrow range of beam parameters.
If the beam pulse is too far off centre, or if the beam size is not correct, or if the beam density is too high, the target can be seriously damaged.
We will start with a brief introduction to high-power targets and then move to a discussion of what can go wrong, and what are the risks.
Next we will discuss how to control the beam-related risk, followed by examples from a few different accelerator facilities.
We will finish with a detailed example of the Oak Ridge Spallation Neutron Source target tune up and target protection.


\end{abstract}
\section{Introduction to high-power targets}

In general, high-power targets are needed to create either lots of secondary particles for applications where the interaction cross-sections are low, or to make as many secondary particles as possible when the creation cross-sections are low. Example applications include:
\begin{itemize}
\item neutron spallation targets (LANSCE, ISIS, SNS, J-PARC, PSI);
\item muon production targets (J-PARC, PSI, TRIUMF, ISIS);
\item Isotope Separation On-Line (ISOL) facilities (CERN, TRIUMF, IRIS);
\item material irradiation studies (IFMIF);
\item antiproton production (FNAL);
\item neutrino production (FNAL, J-PARC, CERN).
\end{itemize}

\begin{figure}[htbp]
\includegraphics [width=\textwidth]{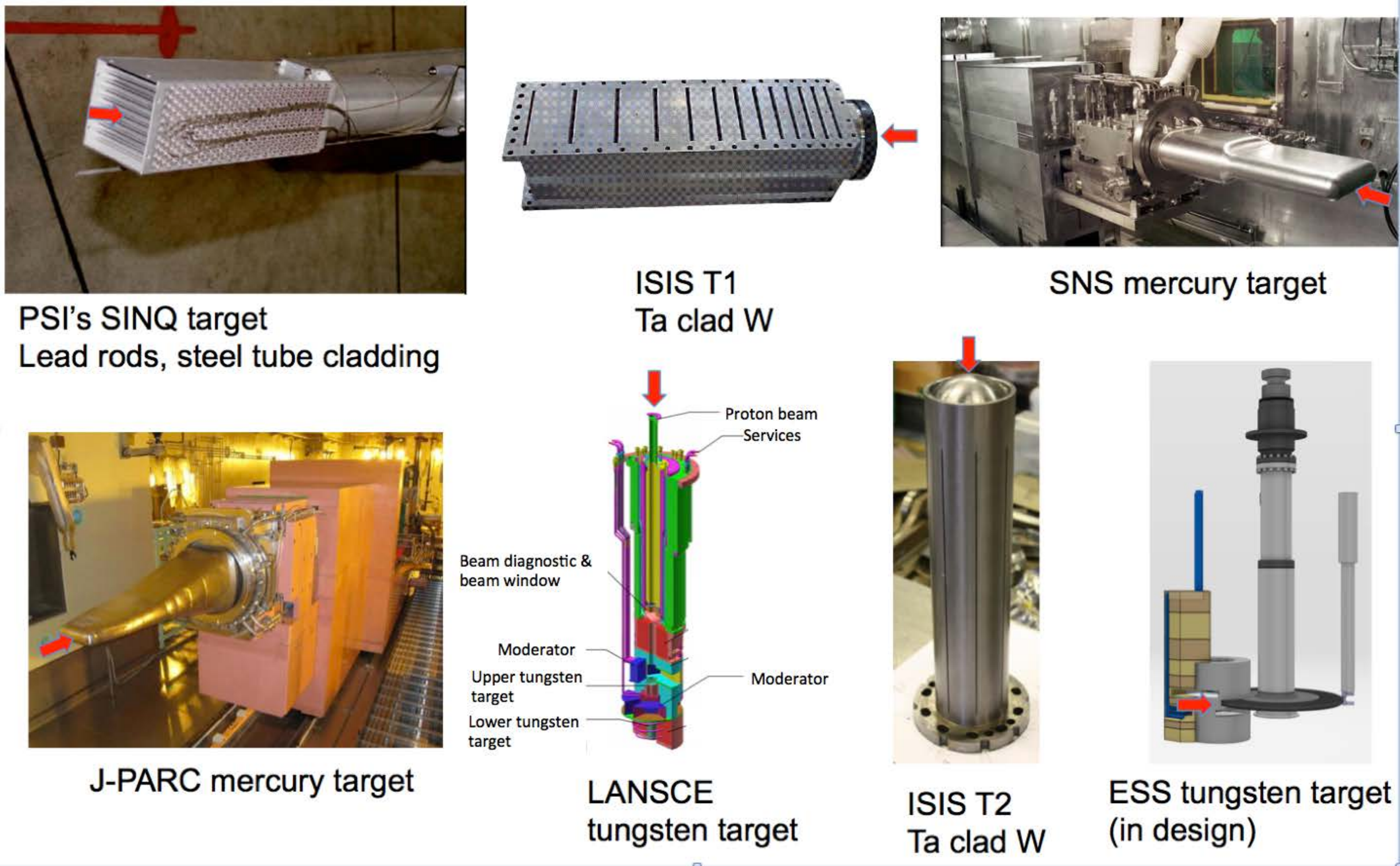}
\caption{\label{fig:spal_tgts} Some example neutron spallation targets. Images reproduced from Refs. \cite{ref:Wagner2011, ref:Reimer2014, ref:Haynes2013, ref:McManamy2009, ref:Werbeck2003}}
\end{figure}

\begin{figure}[htbp]
\includegraphics [width=\textwidth]{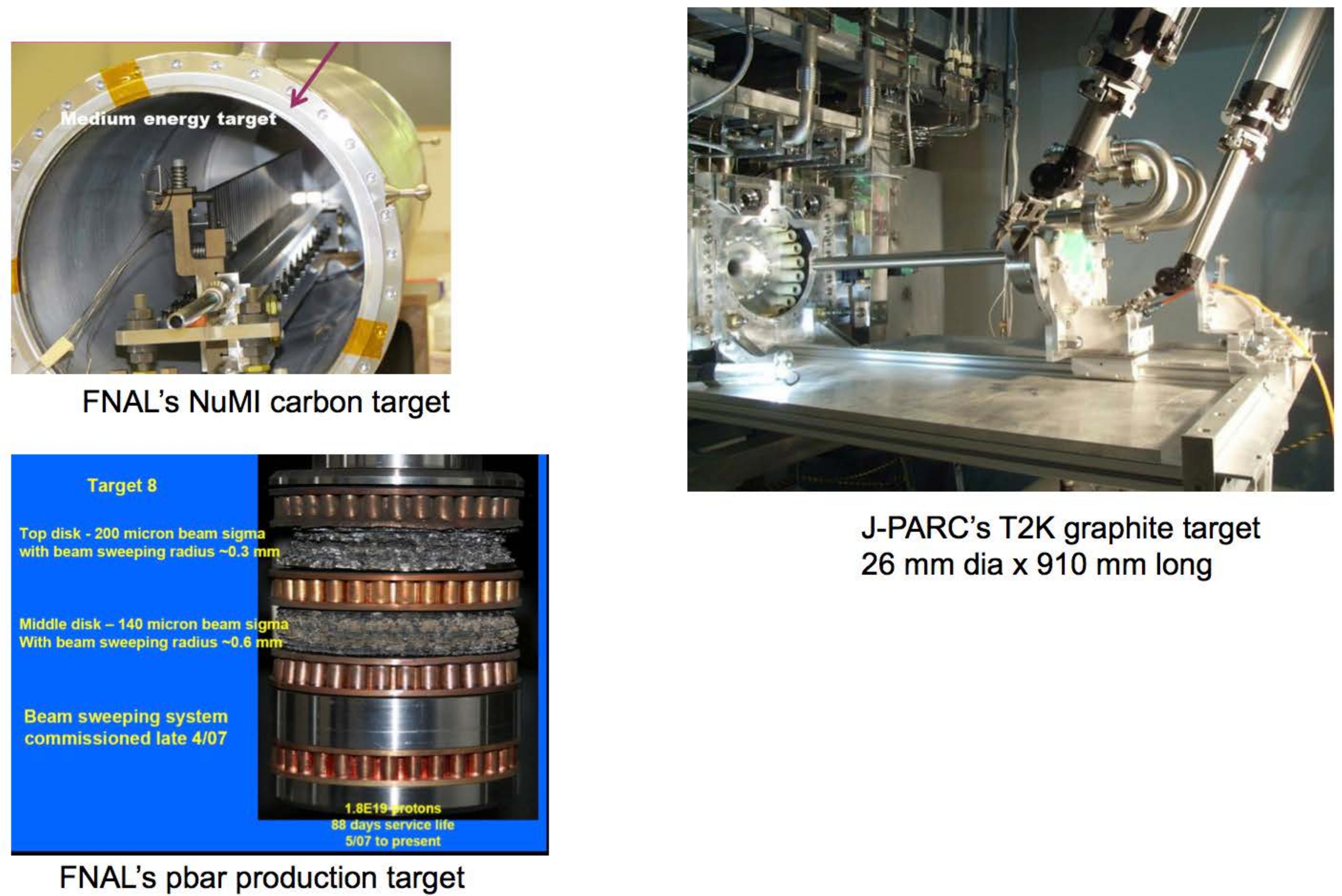}
\caption{\label{fig:nu_tgts} Some example neutrino and pbar targets. Images reproduced from Refs. \cite{ref:Hylen2014, ref:Mokhov2014}}
\end{figure}

\begin{figure}[htbp]
\includegraphics [width=\textwidth]{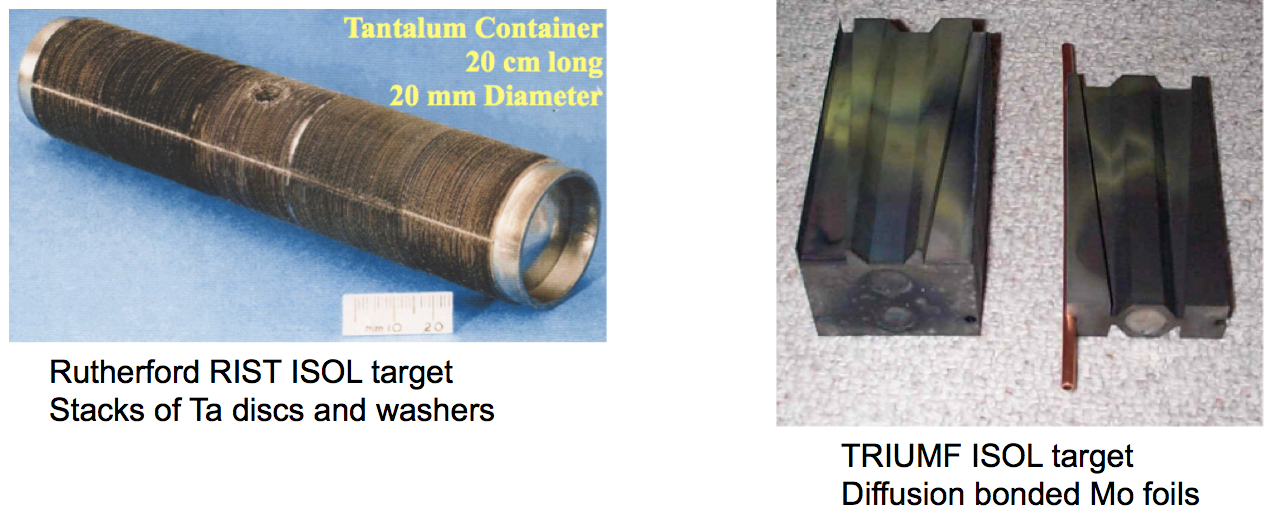}
\caption{\label{fig:isol_tgts} Some example isotope production targets. Images reproduced from Ref. \cite{ref:Bricault2014}}
\end{figure}

High-power targets come in many shapes and sizes, and have many uses. Some example high-power targets are shown in Figs.~\ref{fig:spal_tgts}--\ref{fig:isol_tgts}, and parameters \cite{ref:Haines2009} from some example high-power facilities are shown in Table \ref{tab:hpfac}. High-power targetry is a highly developed and complex field, and there are many technological challenges.
Some of the biggest beam-related challenges are:
\begin{itemize}
\item removing the heat generated mainly by the beam, but also by nuclear decay;
\item mechanical shock due to thermal stress from pulsed beams;
\item radiation damage, including swelling and embrittlement;
\item target handling (including installation and maintenance in a high-radiation environment);
\item corrosive environment;
\item beam parameters---matching target requirements to what the accelerator can deliver.
\end{itemize}

\begin{table}
\caption{Some example high-power target facilities}
\label{tab:hpfac}
\centering
\begin{footnotesize}
\begin{tabular}{p{0.15\textwidth}  p{0.08\textwidth} p{0.07\textwidth} p{0.07\textwidth}  p{0.07\textwidth} p{0.07\textwidth} p{0.07\textwidth}  p{0.07\textwidth} p{0.07\textwidth}}
\hline \hline
\textbf{Facility} & \textbf{Status} & \textbf{Target material} & \textbf{Beam pulse duration} & \textbf{Rep rate} & \textbf{Proton energy} & \textbf{Time avg beam power}& \textbf{Peak time ave power density} & \textbf{Peak energy density}\\
 &  &  & \bf{($\mu$s)} & \bf{(Hz)} & \bf{(GeV)} & \bf{(MW)} & \bf{(GW/m$^3$)} & \bf{(MJ/m$^3$)} \\
\hline
ISIS & Operating & W & 0.4 & 50 & 0.8 & 0.16 & 0.25 & 5 \\
LANSCE--Lujan & Operating & W & 0.3 & 20 & 0.8 & 0.16 & 0.5 & 25 \\
NuMI & Operating & C & 8.6 & 0.53 & 120 & 0.4 & 0.32 & 600 \\
SINQ/Solid target & Operating & Pb--SS clad & CW &  & 0.57 & 1 & 1 & NA \\
SINQ/MEGAPIE & Completed & Pb--Bi & CW &  & 0.57 & 1 & 1 & NA \\
JSNS & Operating & Hg &  & 25 & 3 & 1 & 0.63 & 25 \\
SNS & Operating & Hg & 0.7 & 60 & 1 & 2 & 0.8 & 13 \\
\hline
ESS---long pulse & Proposed & Hg & 2000 & 16.7 & 1.3 & 5 & 2.5 & 150 \\
ESS---short pulse & Proposed & Hg & 1.2 & 50 & 1.3 & 5 & 2.5 & 50 \\
EURISOL & Proposed & Hg & 3 & 50 & 2.2 & 4 & 100 & 2000 \\
IFMIF & Proposed & Li & CW &  & 0.04 (D$_2$) & 10 & 100 & NA \\
LANSCE--MTS & Proposed & Pb--Bi/W & 1000 & 120 & 0.8 & 0.8 & 2.4 & 20 \\
US Neutrino Factory & Proposed & Hg or C & 0.003 & 15 & 24 & 1 & 3.8 & 1080 \\
AUSTRON & Proposed & W & 1 & 10 & 1.6 & 0.5 &  &  \\
\hline \hline
\end{tabular}
\end{footnotesize}
\end{table}

\section{Target environment}

\begin{figure}[htbp]
\begin{center}
\includegraphics [width=0.5\textwidth]{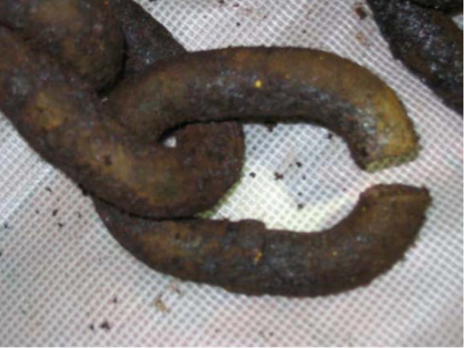}
\caption{\label{fig:broken_chain} High-strength broken steel chain from hydrogen embrittlement caused by acidic vapour/condensate from air ionization (MiniBooNE 25 m absorber). Figure reproduced from Ref. \cite{ref:Hurh2007}.}
\end{center}
\end{figure}

The target environment is often rough vacuum or helium.
Vacuum is good because it does not interfere with the beams (primary or secondary).
Helium atmosphere is good because it helps cool components while minimizing beam scattering; however, impurities in helium can lead to corrosion of components.
Beam transport lines leading up to the target need good vacuum to minimize beam loss due to scattering, so high-power targets often involve windows to separate the beam line vacuum from the target environment.
Safety separation must also be considered (usually required for a liquid metal target).
A partial pressure of air in the target environment can result in the formation of nitric acid, which can cause stress corrosion cracking in high-strength steel.
Nitric acid can also cause vacuum leaks in thin bellows exposed to stray beam with air outside.
For example, IPNS had a target bolt fail from stress corrosion cracking in a nominal helium atmosphere with air impurities when high-strength steel was substituted for stainless steel.
Also, ISIS observed corrosion around the target/reflector assemblies and consequently limited impurities to below a couple of percent in helium.
And, as shown in Fig.~\ref{fig:broken_chain}, Fermilab had a broken chain due to acidic vapour/condensate from air ionization.

\section{Beam windows}

\begin{figure}[htb]
\begin{center}
\begin{tabular}{cc}
    \includegraphics[width=.45\textwidth]{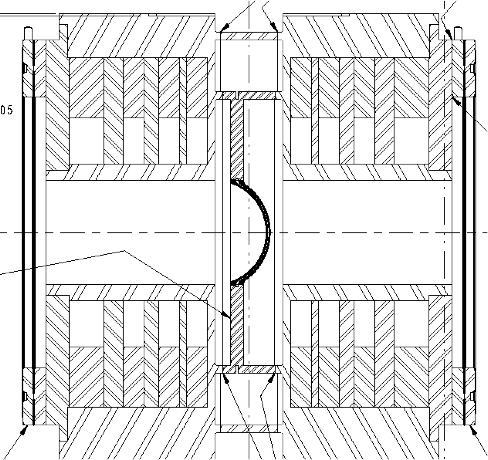} &
    \includegraphics[width=.45\textwidth]{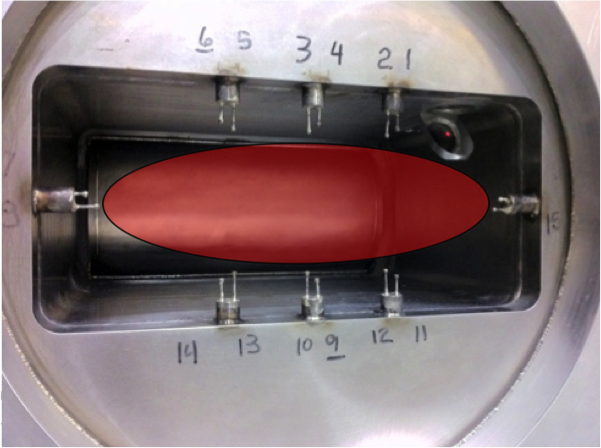} \\
\end{tabular}
\caption{\label{fig:window} Side view and end view of the SNS proton beam window, located about 2 m upstream of the neutron spallation target.  The end view also shows the eight thermocouple halo monitors; the red ellipse indicates the approximate beam size. Images reproduced from Refs. \cite{ref:McManamy2009, ref:McManamy2009b}.}
\end{center}
\end{figure}

Just like the target, beam windows are also challenged by high-power beams, due to heating, thermal stress, radiation damage, etc.
An example beam window is shown in Fig.~\ref{fig:window}. This window is used to separate the SNS beam line vacuum of approximately $1 \times 10^{-7}$~Torr from the target environment, which is 1 atm He.
The window is made of water-cooled Inconel.

\section{Beam parameters}

The highest-power targets push the edge of achievable technology, and to live on the edge requires a very narrow range of beam parameters to avoid overpowering the target.
High-power targets require the nominal beam distribution to be in the correct location, without exceeding the maximum beam current or the maximum beam density.
The most important beam parameters are:
\begin{itemize}
\item beam position;
\item beam size and shape (i.e. distribution);
\item beam energy;
\item beam current;
\item beam pulse length, repetition rate, energy per pulse (short pulses cause a pressure pulse).
\end{itemize}

\begin{figure}[htbp]
\begin{center}
\includegraphics [width=0.5\textwidth]{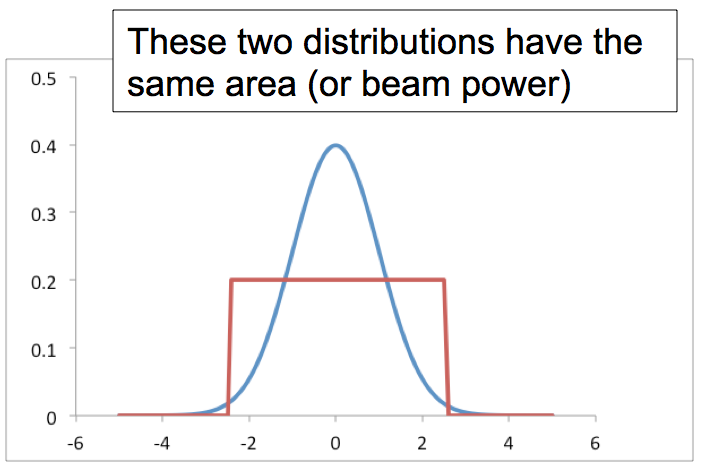}
\caption{\label{fig:beam_dist} Two possible beam distributions. One is rectangular and the other is Gaussian. They both have the same area (or beam power), but the Gaussian distribution has twice the beam density.}
\end{center}
\end{figure}

High-power targets often prefer flat or uniform beam distributions because they minimize the beam density and thereby minimize the density of energy deposited in the target.
However, unless special measures are taken in the accelerator/beam delivery systems, the beam distribution will usually be nearly Gaussian.
A possible target protection requirement could then be to monitor the beam distribution and to shut off the beam if the distribution exceeds requirements.
For example, Fig.~\ref{fig:beam_dist} shows two possible beam distributions. One is rectangular and the other is Gaussian. They both have the same area (or beam power), but the Gaussian distribution has twice the peak beam density.

\subsection{Methods to flatten the beam distribution}

\begin{figure}[h]
\begin{tabular}{ccc}
    \includegraphics[width=.3\textwidth]{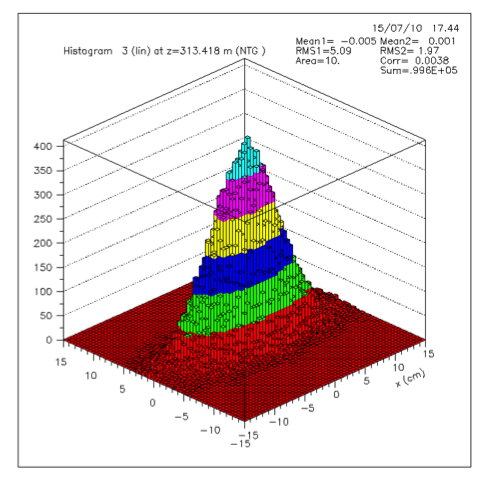} &
    \includegraphics[width=.3\textwidth]{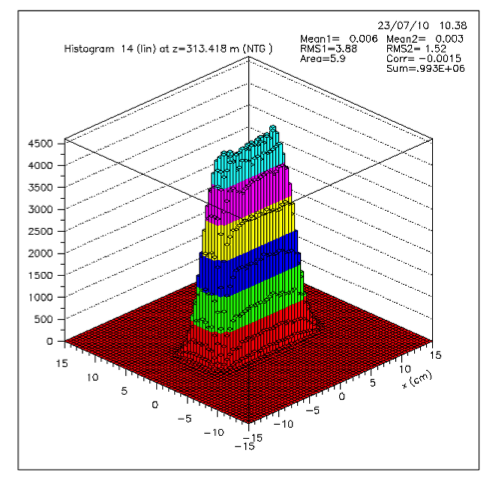} &
    \includegraphics[width=.3\textwidth]{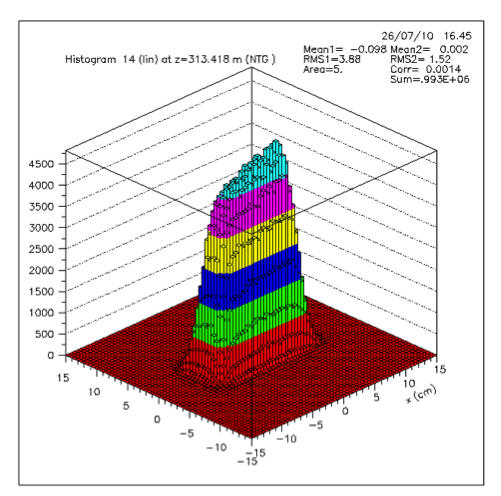} \\
\end{tabular}
\caption{\label{fig:oct} Simulations of beam distributions on the J-PARC neutron production target. Left: no octupoles. Centre: with octupoles. Right: with octupoles and beam off centre at octupoles.  Figures reproduced from Refs. \cite{ref:Meigo2014a,ref:Meigo2014b}.}
\end{figure}

\begin{figure}[h]
\begin{tabular}{c}
    \includegraphics[width=\textwidth]{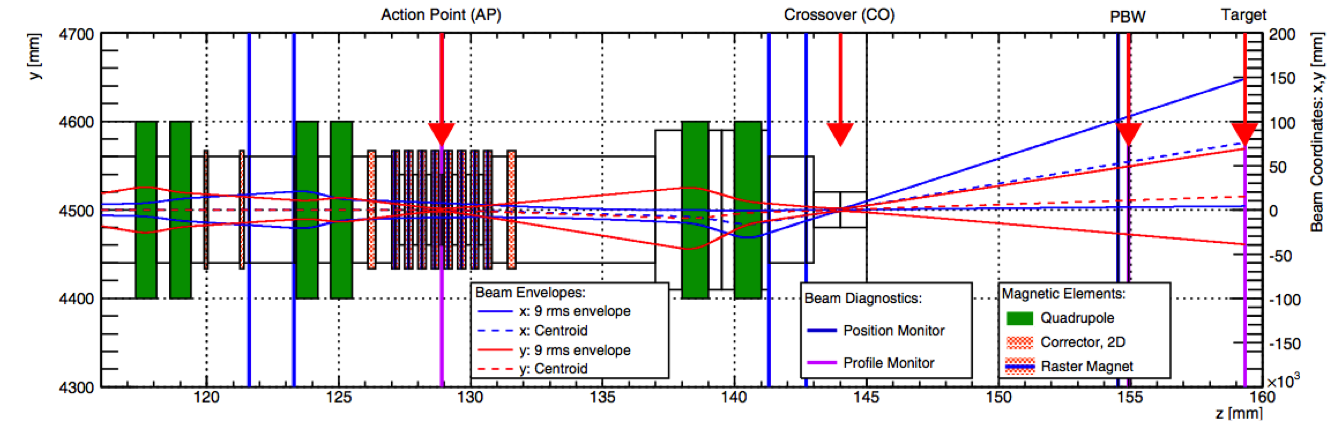} \\
    \includegraphics[width=.4\textwidth]{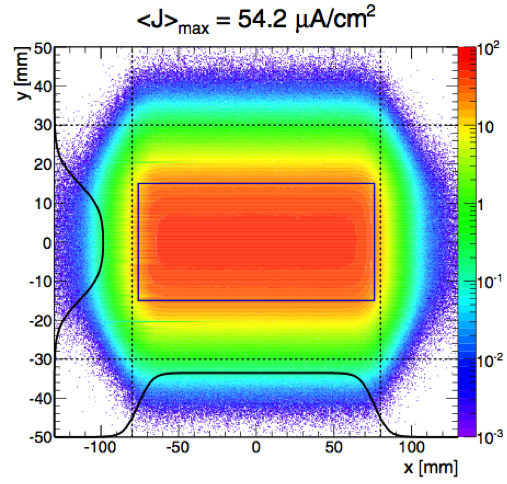}
\end{tabular}
\caption{\label{fig:raster} The ESS design to achieve the desired beam distribution on the target by rastering the beam. Top: the beam transport line just upstream of the target. Bottom: simulation of the achieved beam distribution. Figures reproduced from Ref. \cite{ref:Thomsen2013}.}
\end{figure}

Depending on the application, there are several ways to flatten the beam distribution, to make it more favourable for the target by decreasing the peak density.
For storage rings and synchrotrons that employ multiturn injection (e.g. LANSCE, SNS, J-PARC), injection painting can be an easy way to control the distribution.
Another method is to employ multipole magnets (e.g. octupoles) in the beam transport line to target.
This method was recently implemented at J-PARC, as shown in Fig.~\ref{fig:oct}.
Note that the beam must be well centred in the octupoles for proper flattening, since off-centre beams will produce skewed distributions.
Another method is a rastering system, that `draws' the desired profile by quickly moving a small beam spot on the surface of the target.
This is the method planned for the ESS, as shown in Fig.~\ref{fig:raster}.
All these methods rely on magnets or pulsed power supplies to achieve the desired beam profile.
This highlights possible equipment that would be important to include as inputs to the machine protection system (MPS), and also possible beam monitoring systems, such as beam distribution monitors, that could also be part of the MPS interlock system.

\section{What can go wrong?}

In this section, we briefly describe some examples that highlight the importance of MPSs for high-power target protection.

\begin{figure}[h]
\begin{center}
\includegraphics [width=0.8\textwidth]{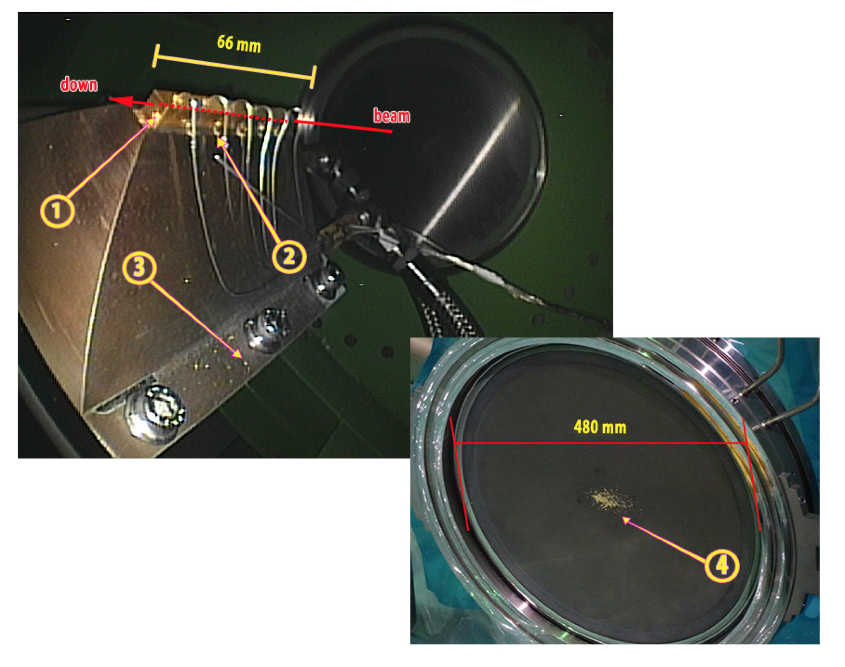}
\caption{\label{fig:jparc_accident} Photographs of the gold target in the J-PARC hadron hall, melted by a proton beam that was accidentally extracted over 5 ms. Figure reproduced from Ref. \cite{ref:jparcaccident}.}
\end{center}
\end{figure}

In May 2013, a gold target (shown in Fig.~\ref{fig:jparc_accident}) melted in the Hadron Experimental Facility at J-PARC and radioactive material was released from the building \cite{ref:jparcaccident}.
The root cause was traced to a quadrupole magnet power supply in the main ring, which caused the beam to be spilled in 5~ms instead of the nominal 2~s.
Due to this accident all beam operations were terminated for about 8 months and beam operations to the hadron hall were stopped for more than one year.

\begin{figure}[h]
\begin{center}
\includegraphics [width=0.5\textwidth]{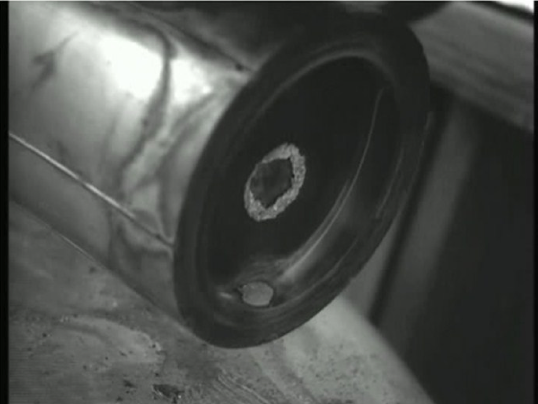}
\caption{\label{fig:isis} A photograph of a failed target from the ISIS second target station. The tantalum cladding on the tungsten target failed.  Figure reproduced from Ref. \cite{ref:ISISTGT2}.}
\end{center}
\end{figure}

\begin{figure}[h]
\begin{center}
\includegraphics [width=0.8\textwidth]{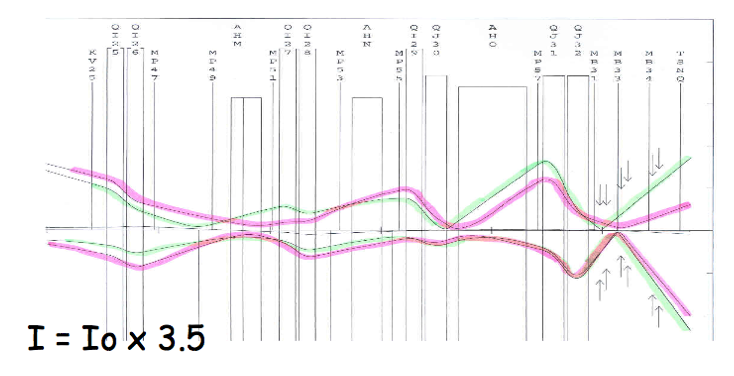}
\caption{\label{fig:VIMOS} The beam transport line leading to the PSI neutron production target, showing the beam sizes for the two different sets of parameters (one for the muon target inserted, the other for the muon target retracted). Figure reproduced from Ref. \cite{ref:Thomsen2007b}.}
\end{center}
\end{figure}

In January 2010, the neutron production target (shown in Fig.~\ref{fig:isis}) failed at the ISIS second target station \cite{ref:ISISTGT2}.
The root cause was traced to a combination of the high-density beam profile and a water leak.
The water disassociated in the high-temperature environment and then the oxygen attacked the grain boundaries on the tantalum cladding.
The cladding essentially fell apart, leaving a hole in the target.
In 2004 the beam density on the PSI neutron production target increased to 3.5 times  the nominal value \cite{ref:Thomsen2007, ref:Thomsen2007b} because the transport line (shown in Fig.~\ref{fig:VIMOS}) was accidentally set up for the case of the muon target being inserted, but actually the muon target was not inserted.
The over-density condition was caught by a newly installed beam distribution monitor (VIMOS) before any serious damage could occur.

We also need to be concerned with beam windows.
They are often part of the target system, and can also be damaged by the beam.
High-power beam windows must be cooled, usually with  water.
Windows can fail due to over-focused or off-centre beams.
Radioactive water can be spilled into the beam line vacuum and/or the target environment.
In 1996 the beam window at the LAMPF linac beam dump failed \cite{ref:Sommer2003}, probably due to a combination of an unusually small beam size (set up on purpose for an experiment) and a weld joint where a thermocouple was attached to the air-side surface of the water-cooled window.

%\FloatBarrier
\section{Calculating the beam density}

To demonstrate how the beam density depends on various beam parameters, we will consider several examples.

\subsection{Example 1: rectangular DC beam}

\begin{figure}[htbp]
\begin{center}
\includegraphics [width=0.3\textwidth]{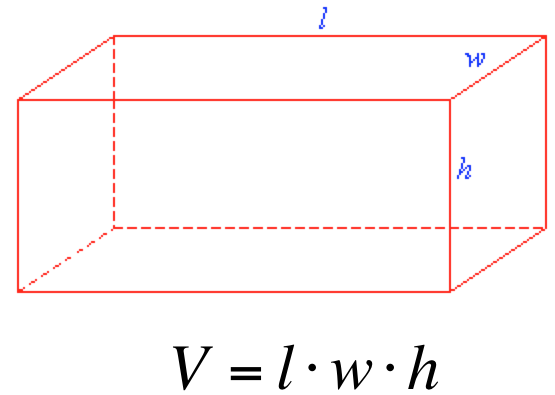}
\caption{\label{fig:rect_beam} A beam pulse with a rectangular cross-section}
\end{center}
\end{figure}

Assume a 2~MW DC beam of 1~GeV protons with rectangular cross-section  2 cm $\times$ 2 cm.
Assume that it stops 0.75 m into the target. Also assume uniform energy deposition and no scattering.
The energy deposition will be as depicted in Fig.~\ref{fig:rect_beam}.
The 1-s power density is then 
$$(2\, \mathrm{ MW})/(2\, \mathrm{ cm})/(2\, \mathrm{ cm})/(0.75\, \mathrm{ m}) = 6.67\, \mathrm{GW/m}^3.$$
Now assume that if the beam is too far off centre it must be shut off before it can deposit 100 J.
What is the acceptable time delay (or MPS response time)?
The beam must be shut off within (100 J)/(2 MJ/s) = 50~$\mu$s.

\subsection{Example 2: rectangular pulsed beam}

Consider a pulsed version of the DC beam in Example 1. Assume that the beam is divided into 10 pulses per second, and the duration of each pulse is 1.5~$\mu$s.
\begin{itemize}
\item The 1-s power density is still (2 MW)/(2 cm)/(2 cm)/(0.75 m) = 6.67 GW/m$^3$.
\item The energy per pulse = (2 MW)/(10 Hz) = 0.2 MJ.
\item The energy density per pulse in target = (0.2 MJ)/(2 cm)/(2 cm)/(0.75 m) = 667 MJ/m$^3$.
\item The peak power density = (667 MJ/m$^3$)/(1.5 $\mu$s) = $4.45 \times 10^{14}$ W/m$^3$.
\end{itemize}

Each beam pulse contains more than 100~J of energy, so even one off-centre pulse will exceed the 100~J limit in Example~1.
With 1.5 $\mu$s beam pulses, it is not practical to shut off the beam mid-pulse.
We can only prevent the next pulse from occurring, and there is a fair amount of time to do that---about 0.1~s.

\subsection{Example 3: pulsed Gaussian beam}

\begin{figure}[htbp]
\begin{center}
\includegraphics [width=0.4\textwidth]{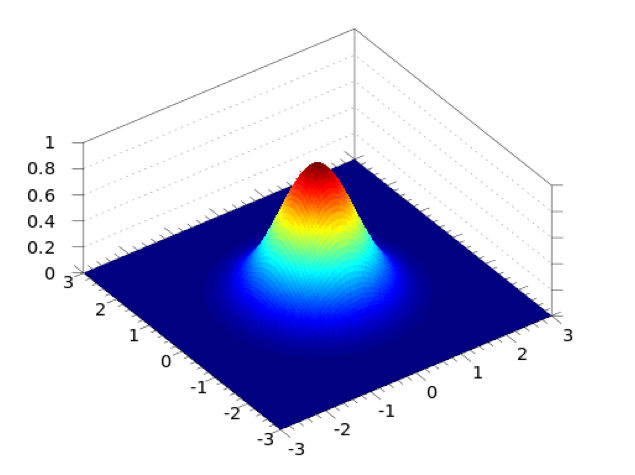}
\caption{\label{fig:gaussian} A two-dimensional Gaussian beam profile}
\end{center}
\end{figure}

Consider a Gaussian version of the pulsed beam in Example 2.
Assume that the Gaussian shape is characterized by $\sigma_x = \sigma_y = 0.5$~cm, as shown in Fig.~\ref{fig:gaussian}.
The functional form of the Gaussian is

$$f(x,y) = A {\rm exp}\left({ -\left(\frac{(x-x_0)^2}{2 \sigma_x^2} +  \frac{(y-y_0)^2}{2 \sigma_y^2} \right)  }\right).$$
The volume under this two-dimensional Gaussian is
$$V = \int_{-\infty}^\infty {\rm d}x \int_{-\infty}^\infty {\rm d}y f(x,y) = 2 \pi A \sigma_x \sigma_y$$
and the 1-s peak power density is
$$(2\, \mathrm{ MW})/(2\pi)/(0.5\, \mathrm{ cm})/(0.5\, \mathrm{ cm})/(0.75\, \mathrm{ m}) = 17\, \mathrm{GW/m}^3,$$
which is 2.25 times greater than the rectangular distribution case.
The per-pulse peak energy density and peak power density also increase by a factor of 2.5.

\section{Target protection}

We will consider only the beam-related protection.
The targets themselves will have their own separate non-beam-related protection systems to monitor coolant flow, over temperature, etc.
The main beam parameters of concern are density (sometimes derived from a beam profile measurement), beam size, beam position, and beam current (from which beam power can be calculated).
Beam-related target protection involves control over these parameters, monitoring these parameters, and turning off the beam if these parameters move outside allowable limits.
Types of beam-related target protection include:
\begin{itemize}
\item rapid and automatic beam turn off for off-normal beams;
\item locking down equipment and operating control parameters to prevent accidental changes;
\item alarms that alert operators when certain parameters move outside of pre-defined limits;
\item collimators to partially intercept off-normal beams;
\item beam transport designs that avoid high sensitivity---e.g. do not want to live on the edge, do not want to be very sensitive to small magnet changes;
\item target designs that can tolerate off-normal beam for a short period of time (e.g. most SNS equipment is designed such that it can withstand two full-power off-normal beam pulses).
\end{itemize}

\begin{figure}[htbp]
\begin{center}
\includegraphics [width=0.8\textwidth]{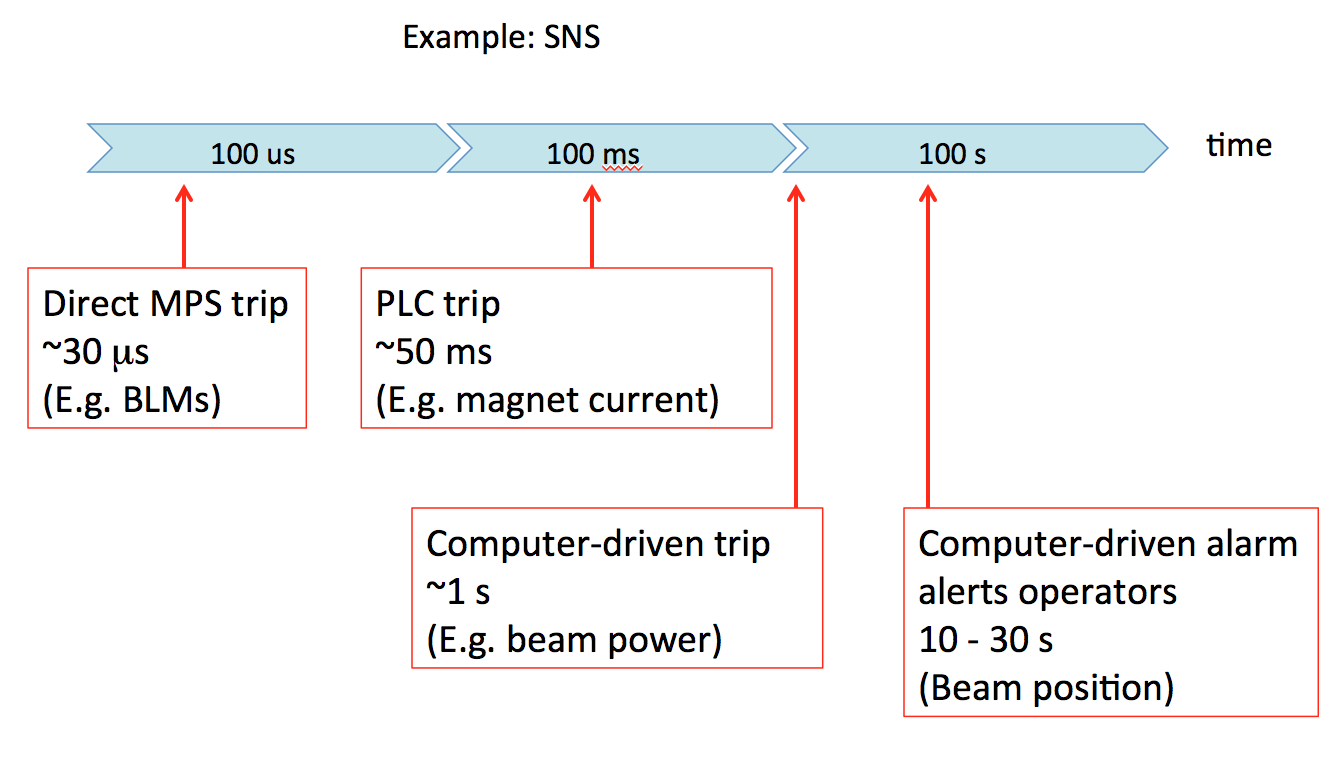}
\caption{\label{fig:time_to_trip} Some typical turn-off times for the SNS case}
\end{center}
\end{figure}

\subsection{Protection by rapid beam turn off}

The time it takes to automatically  turn off the beam varies depending on the method used.
Figure~\ref{fig:time_to_trip} shows some typical turn-off times for the SNS case.

The fastest method to turn off the beam is through the MPS.
After all, that is the purpose of the MPS.
A typical time to turn off the beam at SNS is 20 to 30~$\mu$s for trips that originate in the linac, for example if a beam loss monitor exceeds a pre-determined threshold.
Trips that originate further downstream, in the ring or in the ring-to-target beam transport, can take a few $\mu$s longer.
An interesting fine point is that if the trip occurs upstream of or within the accumulator ring, the MPS can stop beam injection, but the beam that has already been injected into the
ring will be sent to the target.
The beam intensity can be anywhere from small to full intensity.
If the trip occurs downstream of the ring, it can only prevent the next pulse from beginning the accumulation process.

For example, at SNS, PLCs monitor  magnet current transformers on the power supplies for the last few magnets in the transport line to the target.
If the magnet current strays outside pre-determined limits, the beam will be automatically turned off.
The  time for a PLC trip to turn off the beam is about 50~ms.
As an additional precaution,  PLC trips also turn off the high voltage to the ion source and turn off the high voltage to the first few klystrons.

\begin{figure}[htbp]
\begin{center}
\includegraphics [width=\textwidth]{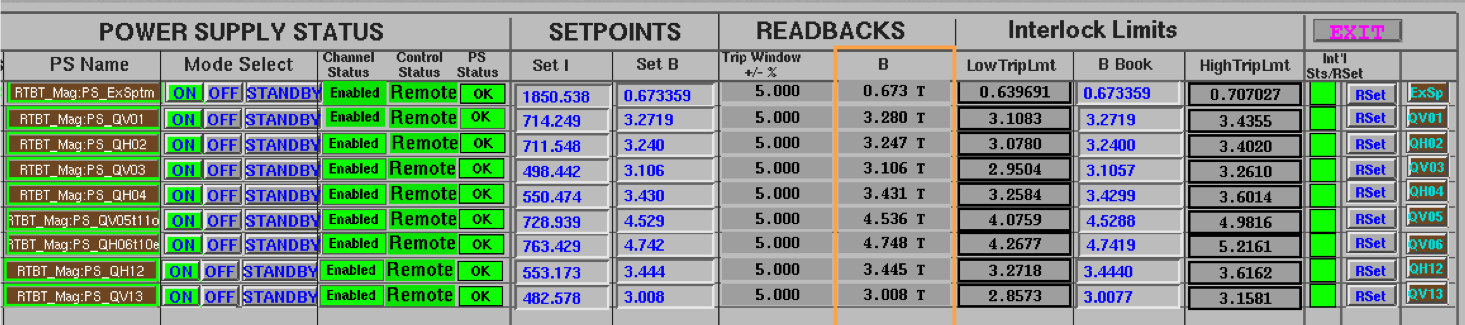}
\caption{\label{fig:ps_curr_limits} Some of the power supplies in the SNS beam transport from the ring to the target, and their trip limits that are monitored by the control computer.}
\end{center}
\end{figure}

The third-fastest method is faults detected by the control computers.
The computers interface to the MPS to turn off the beam.
For example, at SNS, computers monitor all the magnet power supply currents in the beam transport to the target (as read by the power supply controller---not by separate current transformers).
The estimated time to turn off the beam is about 1~s.
For example, Fig.~\ref{fig:ps_curr_limits} shows a partial list of these magnets and their high- and low-trip limits.

The fourth-fastest method is the alarm system, which alerts the operators to off-normal conditions.
The alarms are generated by the control system, and require operator action to turn off the beam or to correct the off-normal condition.
A strength of this method is that the control system can monitor a huge number of parameters, and also values derived from those parameters.
For example, at SNS, the control system alarms if the beam position on target strays outside of pre-determined limits.
The estimated time to correct the alarm state is 10 to 30~s, or possibly longer, depending on the operator reaction time and the complexity of the alarm.

\begin{figure}[htbp]
\begin{center}
\includegraphics [width=0.7\textwidth]{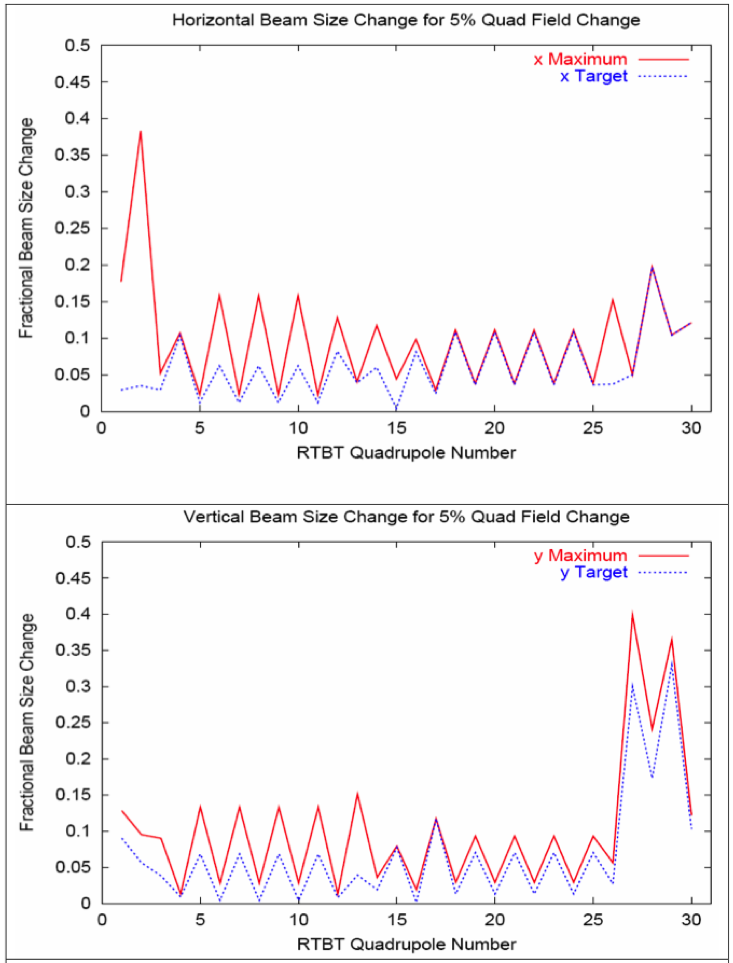}
\caption{\label{fig:sns_quad_sensitivity} These plots show, for each quadrupole magnet in the beam line from the ring to the target, how much the beam size on the target changes for a 5\% change in the quadrupole magnet gradient (blue line).  Also shown is the maximum beam size fractional change along the beam line (red line). Figure reproduced from Ref. \cite{ref:Holmes2005}.}
\end{center}
\end{figure}

\begin{figure}[htbp]
\begin{center}
\includegraphics [width=0.7\textwidth]{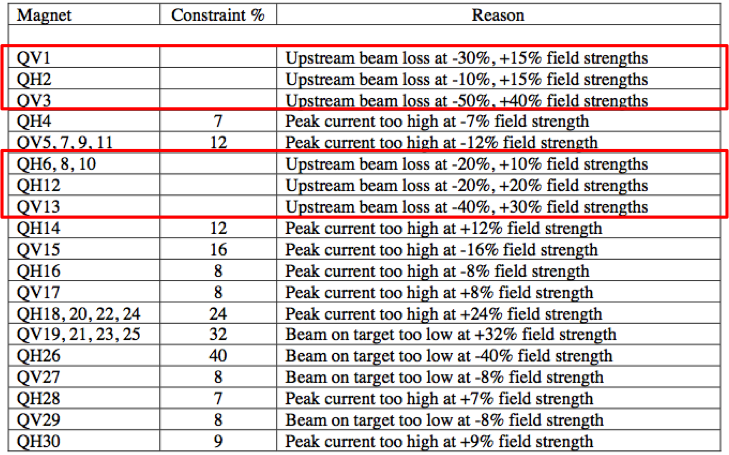}
\caption{\label{fig:sns_quad_sensitivity2} The six of 19 quadrupole magnets that will cause a BLM trip (beam loss greater than 0.1\%) before causing dangerous beam parameters at the target. Figure reproduced from Ref. \cite{ref:Holmes2005}.}
\end{center}
\end{figure}

Beam loss monitors (BLMs) are a standard part of machine protection for any beam transport system, and their purpose is primarily  to protect beam line components.
But BLMs can also be used to protect a target.
The idea is that a quadrupole magnet change that can cause the beam to be too small or too large at the target will sometimes also cause beam loss upstream of the target, because the mis-focused beam could be large enough somewhere along the beam line that the beam tails will strike the beam pipe walls.
BLMs can also protect against deviations from the nominal beam trajectory that could result in a bad position on the target, if the deviation is large enough to cause beam loss.
At SNS, BLMs are located, and thresholds are set, such that some of the quadrupole magnet changes that can result in off-normal beams at the target will trip the BLMs.
Figure~\ref{fig:sns_quad_sensitivity} shows, for each quadrupole magnet in the beam line from the ring to the target, how much the beam size on the target changes for a 5\% change in the quadrupole magnet gradient.
It also shows the greatest amount of change anywhere in the beam line.
Figure~\ref{fig:sns_quad_sensitivity2} shows the six of 19 quadruple magnets that will cause a BLM trip (beam loss greater than 0.1\%) before causing dangerous beam parameters at the target.

\subsection{Protection by equipment lock down}

Another method to protect the target is to lock down certain control parameters.
The idea is to administratively control critical hardware set points to prevent inadvertent changes.
An example of this method is that at SNS the operators set a gate generator at the ion source that limits the possible beam pulse lengths, which limits the beam power on the target.

\subsection{Protection by collimation}

\begin{figure}[htbp]
\begin{center}
\includegraphics [width=0.9\textwidth]{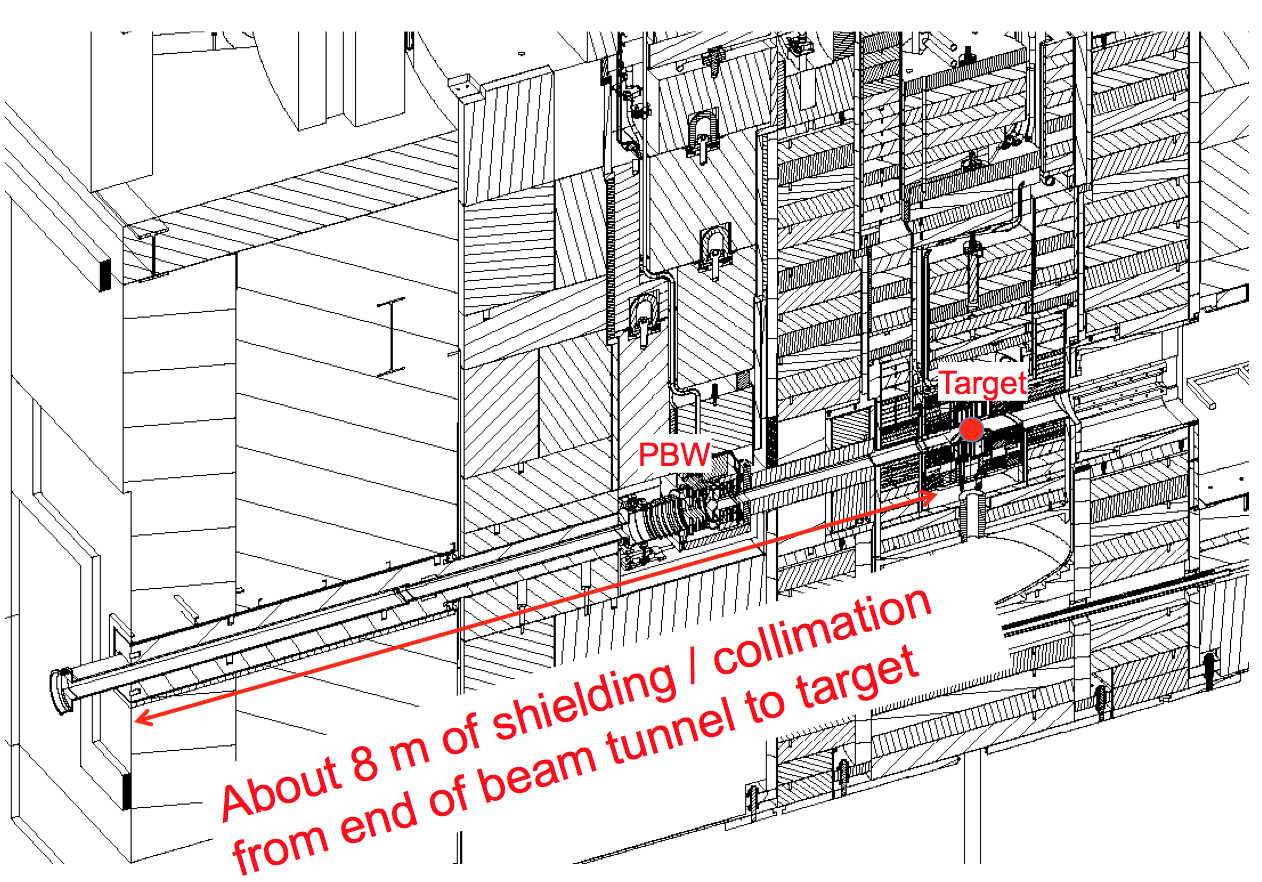}
\caption{\label{fig:tgt_coll} The collimator and shielding just upstream of the SNS neutron production target. The PBW label refers to the Proton Beam Window.}
\end{center}
\end{figure}

\begin{figure}[htbp]
\begin{center}
\includegraphics [width=0.9\textwidth]{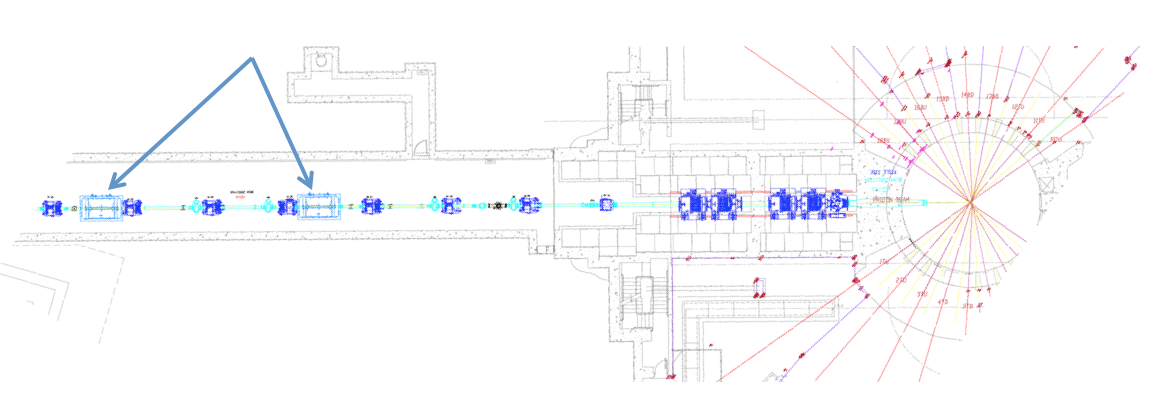}
\caption{\label{fig:rtbt_tgt} The downstream portion of the SNS RTBT beam line. The blue arrows show the positions of the RTBT collimators.}
\end{center}
\end{figure}

\begin{figure}[htbp]
\begin{center}
\includegraphics [width=0.9\textwidth]{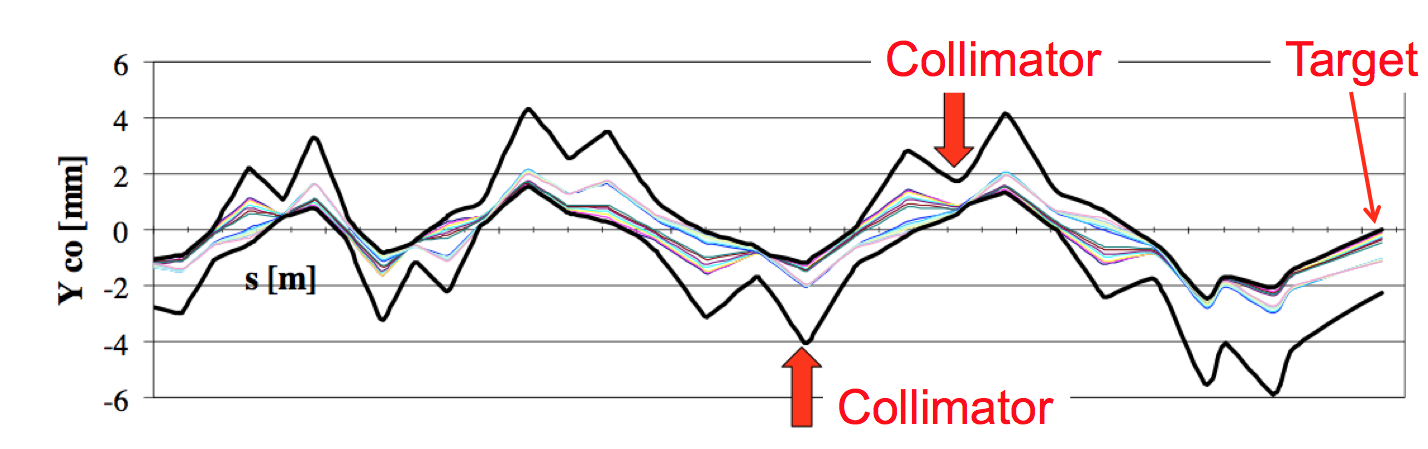}
\caption{\label{fig:rtbt_coll_bnl} Simulations of beam trajectory deviations caused by missed kicks in the ring extraction kicker system. The thin lines indicate trajectory deviations for individual missed kicks, and the thick lines indicate the maximum and minimum trajectory deviations for the case of two simultaneous missed kicks. Figure reproduced from Ref. \cite{Catalan2001}.}
\end{center}
\end{figure}

High beam power facilities often have collimators in the last part of the beam transport leading to the target, to protect against large beam position variations and overly large beam sizes, and to ensure that the beam hits the central region of the target.
For example, SNS has a collimator immediately upstream of the target.
The $27.9 \times 12.7$~cm$^2$ aperture is smaller than the face of the target, but slightly larger than the nominal beam size (90\% of beam must fit within a $20 \times 7$~cm$^2$ rectangle).
SNS also has collimators in the beam transport line that can intercept a portion of the beam that does not receive the nominal kick angle from the extraction kickers, thus protecting the target from some of the mis-kicked beam that would consequently arrive off centre at the target.
Figure~\ref{fig:tgt_coll} shows the collimator that is embedded in the bulk shielding surrounding the SNS target,  Fig.~\ref{fig:rtbt_tgt} shows the positions of the beam line collimators, and Fig.~\ref{fig:rtbt_coll_bnl} shows simulations of the beam trajectory deviations due to extraction kickers that do not fire properly.

\subsection{Protection by interlocking on off-normal measured beam parameters}

This protection category involves continuously monitoring critical beam parameters, and then automatically turning off the beam if the parameters exceed pre-determined thresholds.
Table~\ref{tab:interlock} shows some of the more common beam instrumentation used for this purpose.

\begin{table}
\caption{Some example target interlocks}
\label{tab:interlock}
\centering
\begin{tabular}{p{0.3\textwidth}  p{0.6\textwidth}   }
\hline \hline
\textbf{Interlock} & \textbf{Some accelerators that use this as a target interlock} \\
\hline
Beam loss monitors & Everybody \\
Beam current monitors & Everybody \\
Beam position monitors &  PSI and FNAL have automatic beam centring based on beam position monitor signals \\
Harp profile monitors & ISIS$^{\rm a}$, LANSCE$^{\rm a}$, J-PARC$^{\rm a}$, SNS$^{\rm a}$ \\
Ionization profile monitors & PSI \\
Glowing screen just upstream of target (profile monitor) & PSI \\
Halo thermocouples & SNS \\
Amount of beam intercepted by collimator & PSI \\
\hline \hline
\multicolumn{2}{l}{ $^{\rm a}$ No interlock at present but this is under consideration}
\end{tabular}
\end{table}

\subsection{Protection by beam line design}

It is usually desirable to design the beam transport  line to avoid high sensitivity to beam parameters at the target.
For example, small adjustments to dipole magnets should not cause large position changes on the target, and small adjustments to quadrupole magnets should not cause large beam size changes on the target.
At SNS, one way this design practice was implemented was to require the phase advance from the extraction kickers to the target to be approximately a multiple of $\pi$.
In this way, a missed kick from an extraction kicker will not cause a large position change on the target. This effect is illustrated in Fig.~\ref{fig:rtbt_coll_bnl}.

\section{Some example target protection implementations}

\subsection{J-PARC target protection}

The J-PARC design parameters call for a 1 MW, 3 GeV, 25 Hz proton beam to be delivered to a liquid mercury target.
To protect the target \cite{Kinsho2014}, there is a waveform monitor on each of the eight extraction kickers.
A PLC monitors some of the quadrupole and dipole magnet power supply currents.
A PLC also monitors the beam profile measurement device upstream of the target, and it is possible to connect this to the machine protection system, but presently this is not the case.
A recent addition to the MPS interlock system is a fast beam current monitor to shut off the beam if the beam current exceeds a pre-defined limit.

\subsection{FNAL NuMI target protection}

The NuMI neutrino production target at FNAL now operates with 120 GeV, approximately 400 kW protons, <1~Hz, with an upgrade to 700~kW planned in 2015.
A later upgrade to  >1~MW is planned.
Prior to beam extraction from the main injector (MI), more than 250 different inputs to the beam permit system are checked \cite{Childress2008}.
The beam is sent to the abort dump if inputs are not correct.
The inputs checked include the following: beam position and angle for the MI extraction channel, possible excessive residual MI beam in the kicker beam gap, the extraction kicker status, proper NuMI power supply flat-top values, the beam readiness of the target station and absorbers, all beam loss monitor readings from the previous extraction, and the previous pulse position and trajectory at the target.
The beam delivery system also has an automatic beam steering system to keep the beam centred on the target.
If a beam pulse is >1.5~mm off centre, the beam is automatically turned off.

\subsection{ISIS target protection}

The ISIS facility at RAL operates two targets, TS1 and TS2.
TS1 operates with 800 MeV, 160 kW, 40~Hz protons, and TS2 operates with 800~MeV, 40~kW, 10~Hz protons.
A beam halo monitor (comprising eight equally spaced thermocouples in the penumbra of the beam $\sim$100 mm upstream of the target) indicates mis-focused and mis-steered beams.
If the thermocouple signals exceed pre-determined trip thresholds, the protection system turns off the beam.
The protection system \cite{Adams2014} is capable of turning off the machine within 2~ms and hence inhibits the next beam pulse from the ion source in the 50 Hz synchrotron cycle.
There is also  a steering servo system to accommodate variations in the upstream beam.
The servo and trip system is slow, running at $\sim$2~s.
On TS2 there is also a harp profile monitor which sits permanently in the beam, but this is not interlocked.
A similar harp is planned for TS1 as part of an upgrade project.

\section{SNS target tune up}

\begin{figure}[!htb]
\begin{center}
\includegraphics [width=0.9\textwidth]{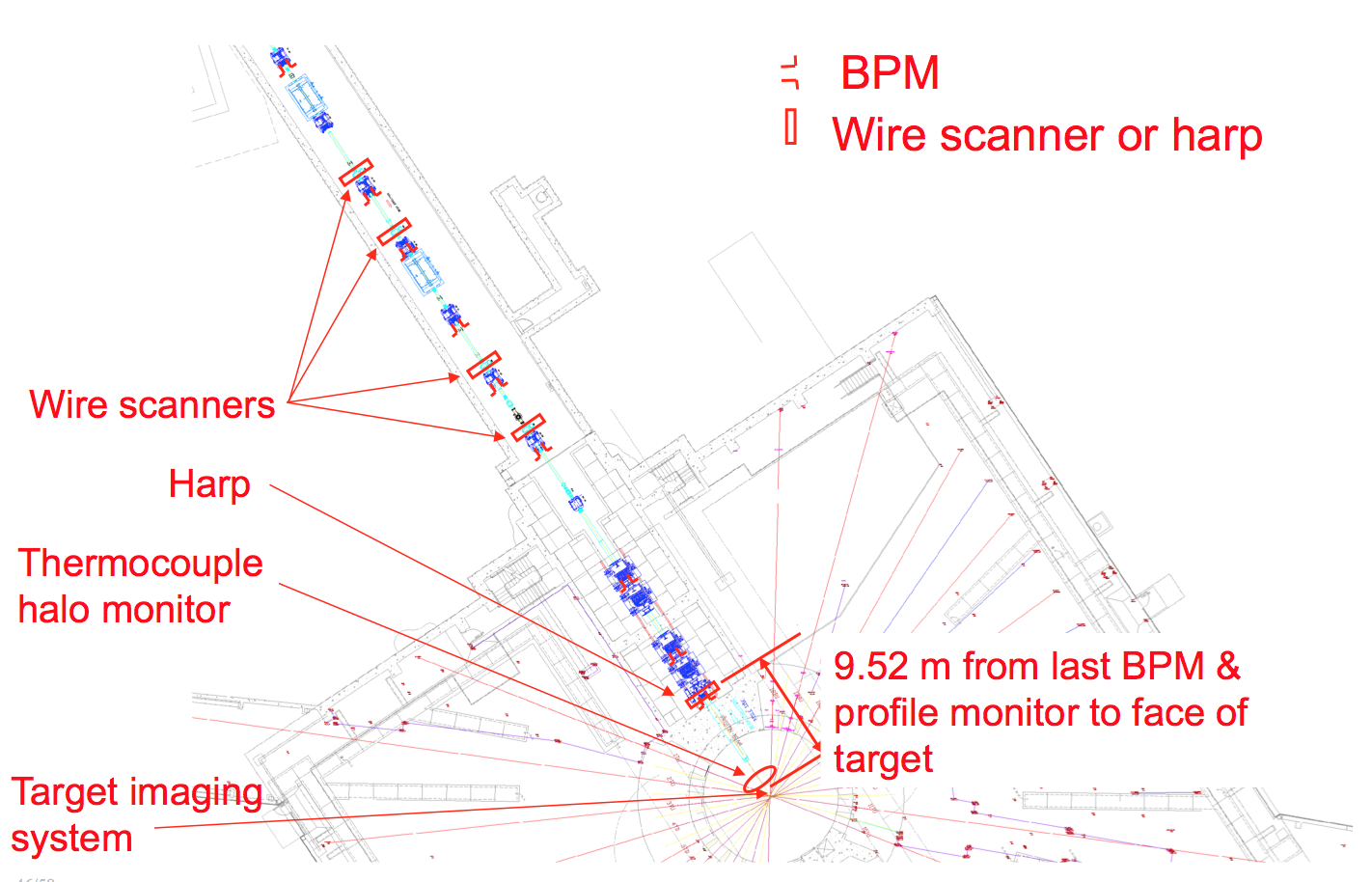}
\caption{\label{fig:rtbt_bi} A drawing of the SNS beam line upstream of the target, showing the locations of the profile and position monitors. Image reproduced from Ref. \cite{ref:Plum2008}.}
\end{center}
\end{figure}

\begin{figure}[!htb]
\begin{center}
\includegraphics [width=0.9\textwidth]{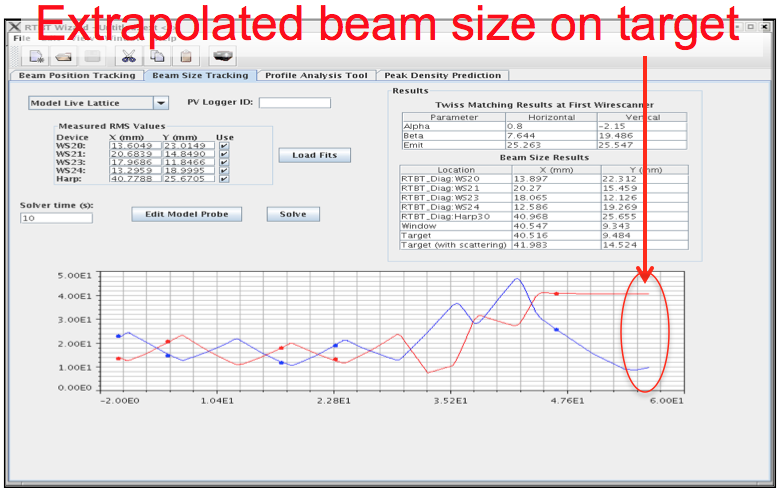}
\caption{\label{fig:rtbt_fit} Screen shot of the application that fits a model to measured beam sizes to extrapolate the beam size at the target. The points show the measured beam sizes and the lines show the model beam sizes.}
\end{center}
\end{figure}

\begin{figure}[!htb]
\begin{center}
\includegraphics [width=0.9\textwidth]{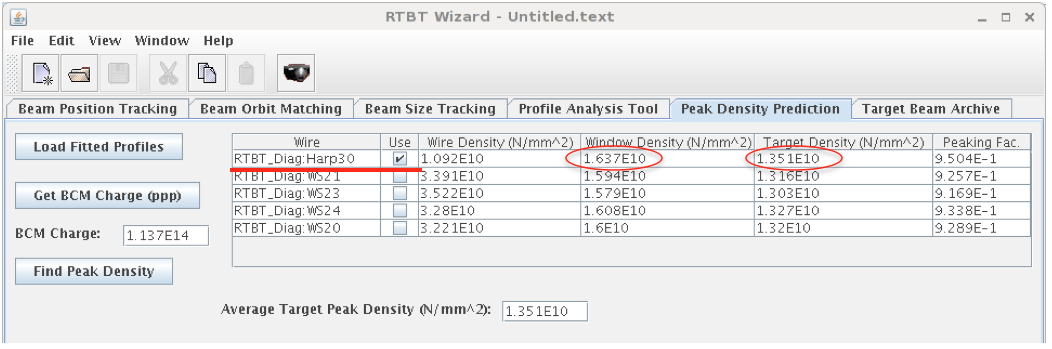}
\caption{\label{fig:rtbt_fit_table} Another screen shot of the application that extrapolates the beam size on the target. The resultant size predictions at the beam window and target are circled.}
\end{center}
\end{figure}

To provide a detailed example of target protection, we will now discuss the SNS case.
We will be referring to the beam instrumentation shown in Fig.~\ref{fig:rtbt_bi}.
The initial beam tuning to the target is performed at low intensity, low power, 1~Hz (<1~kW) beam.
Once the beam position has been adjusted to centre the beam on the target, the beam power is slowly increased to full intensity, but still at 1~Hz ($\sim$23~kW).
The beam intensity is now high enough that an image can be seen on the target imaging system (TIS), which is based on a light-emitting coating on the surface of the target, and hence gives a direct measurement of beam position and beam distribution at the target.
The TIS is used to check the beam centring, and then the beam size is measured not with the TIS (due to unresolved discrepancies with the measurement technique about to be discussed), but with the four wire scanners and one harp located at various positions in the ring to target beam transport (RTBT) beam line, as shown in Fig.~\ref{fig:rtbt_bi}.
The beam size at the target is determined by fitting the measured rms beam sizes with an online model, and varying the simulated beam parameters until the best fit to the measured rms beam sizes is obtained.
The model is then used to extrapolate the beam size at the target.
An example fit is shown in Fig.~\ref{fig:rtbt_fit}, and Fig.~\ref{fig:rtbt_fit_table} shows the corresponding output table with the extrapolated beam size at the target.

The same online model is then used to extrapolate the peak beam density, measured with the harp, to the target.
The critical beam parameters for the target have now been measured.
To meet requirements:
\begin{itemize}
\item the beam must be centred on the target to within 6 mm horizontal and 4 mm vertical;
\item the rms beam size must be <49 mm horizontal and <17 mm vertical;
\item the peak density on target must be less than $2 \times 10^{16}$ protons/m$^2$;
\item the peak density on the proton beam window must be less than $2.9 \times 10^{16}$  protons/m$^2$.
\end{itemize}

\begin{figure}[!htb]
\begin{center}
\includegraphics [width=0.9\textwidth]{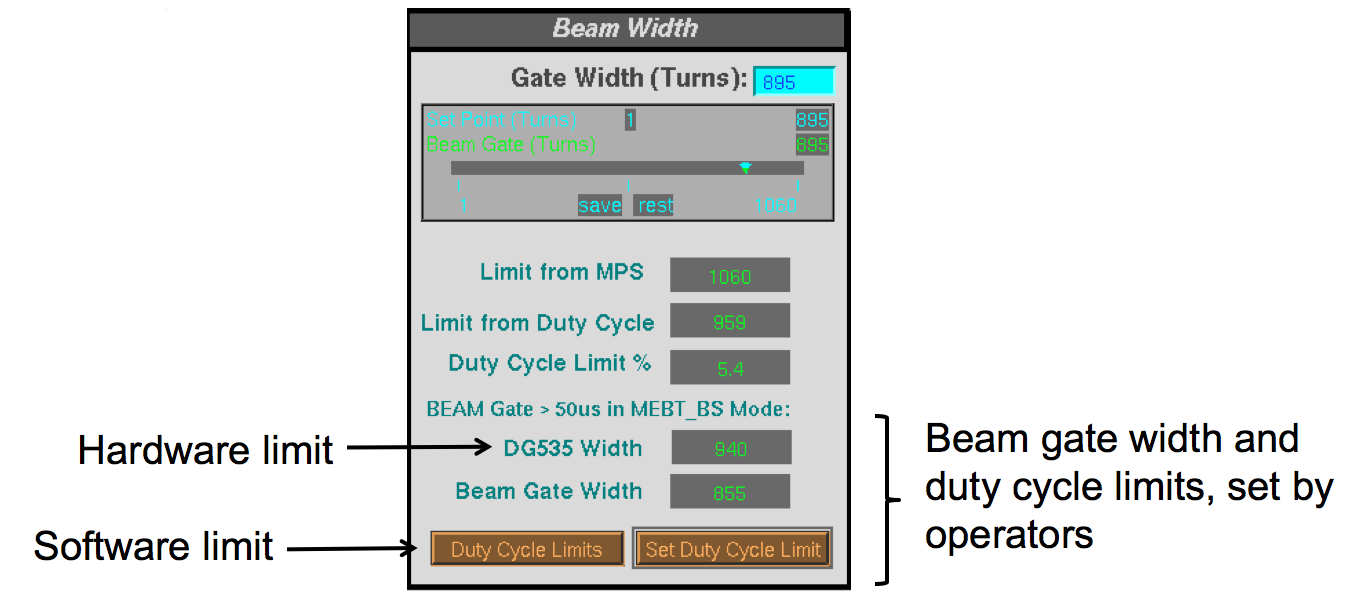}
\caption{\label{fig:beam_gate} A screen shot of the application showing the DG535 pulse generator settings that lock down the maximum possible beam pulse width.}
\end{center}
\end{figure}

Once the measured beam parameters have been determined to meet requirements, the beam delivery system is locked down.
This is required before increasing the beam power above 100~kW.
The lock down is accomplished by engaging the following interlock thresholds.
\begin{itemize}
\item The last five quadrupole magnet power supplies are monitored by a PLC with current limits set to $\pm$7\%.
\item Both large dipoles are monitored by a PLC with current limits set to $\pm$2 A.
\item The last two horizontal dipole correctors and last two vertical dipole correctors are monitored by a PLC with current limits set to $\pm$5 A.
\item The injection kicker waveform monitor system is engaged. If waveforms stray outside of pre-defined windows, the MPS trips the beam.
\item The extraction kicker waveform monitor system is engaged. If waveforms stray outside of pre-defined windows, the MPS trips the beam.
\item All RTBT magnets are monitored by the control computer with current limits set to $\pm$5\% on quadrupole magnets, $\pm$5\% on large dipole magnets, and $\pm$0.5~A on dipole corrector magnets.
\item The beam power limit is set into the control system, and monitored by the computer.
\end{itemize}
The beam pulse length and beam duty factor are also locked down, but in a way that does not cause a beam trip.
It simply prevents the timing system hardware from being set in a way that could exceed a pre-determined beam power.
A screen shot of this system is shown in Fig.~\ref{fig:beam_gate}.

\begin{figure}[!htb]
\begin{center}
\includegraphics [width=0.9\textwidth]{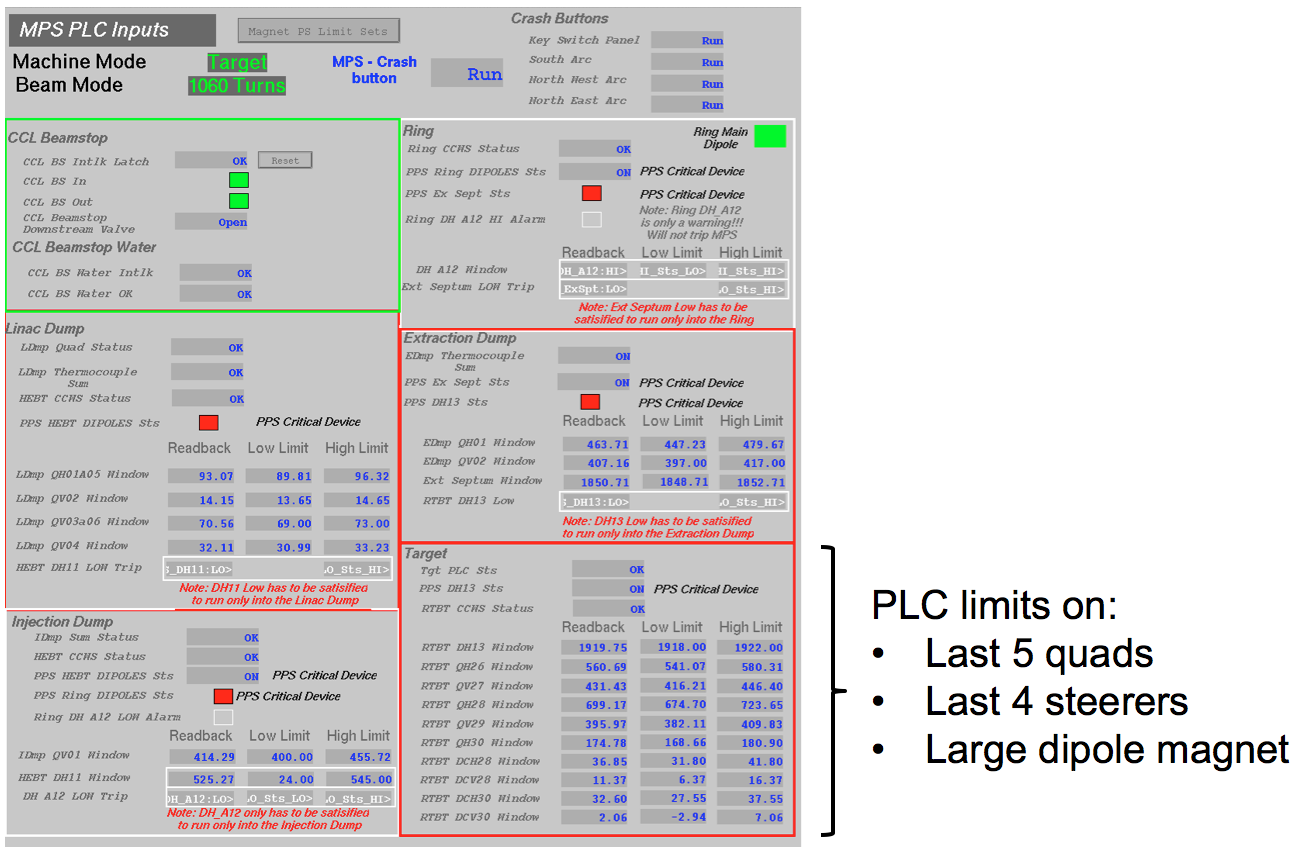}
\caption{\label{fig:mps_plc} A screen shot showing the PLC limits on the last magnets upstream of the target}
\end{center}
\end{figure}

\begin{figure}[!htb]
\begin{center}
\includegraphics [width=0.9\textwidth]{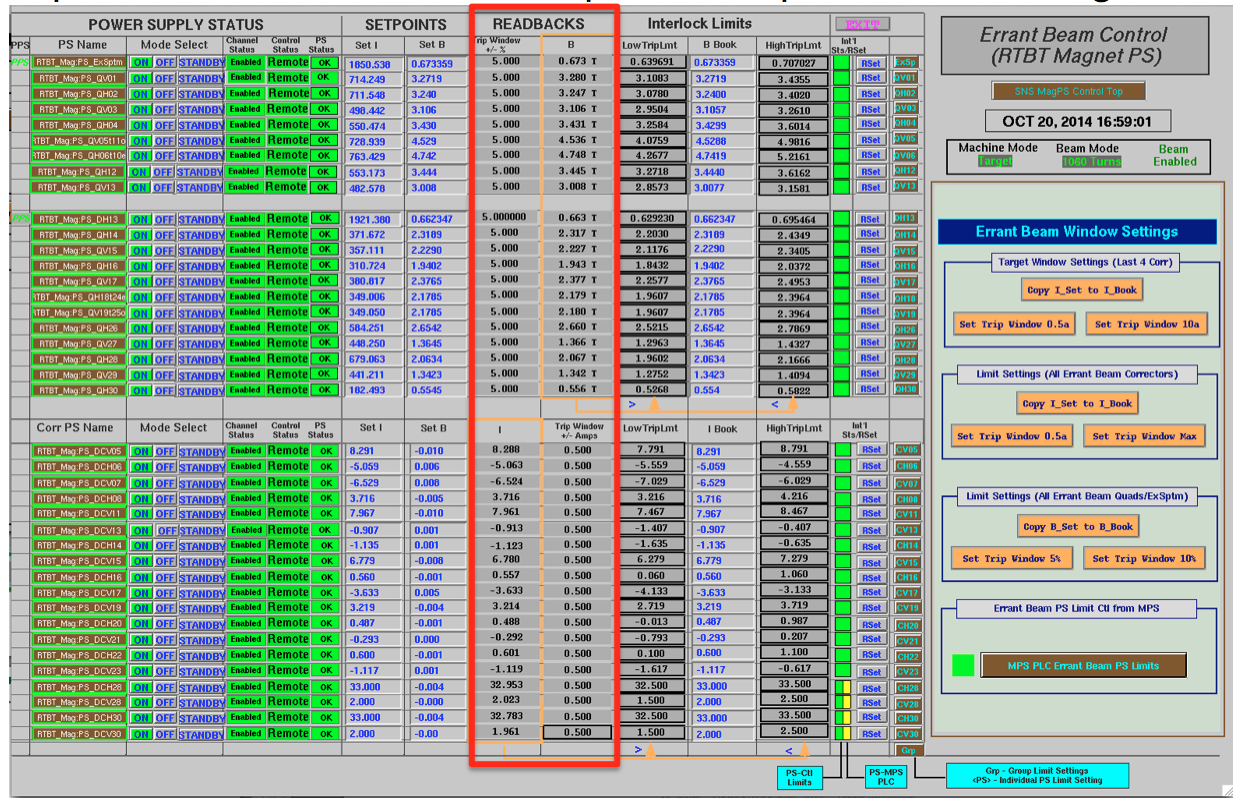}
\caption{\label{fig:mag_lockdown} A screen shot showing the magnets monitored by the control system to lock down the beam transport to the target.}
\end{center}
\end{figure}

Figure~\ref{fig:mps_plc} shows a screen shot of the magnet currents that are monitored by a PLC.
The PLC monitors those magnets that have the biggest impact on beam size and position at the target.
The control system computers also monitor these magnets, as well as many more, as shown in Fig.~\ref{fig:mag_lockdown}, but the PLC is a more robust and reliable system compared to the control computer.

\begin{figure}[!htb]
\begin{center}
\includegraphics [width=0.9\textwidth]{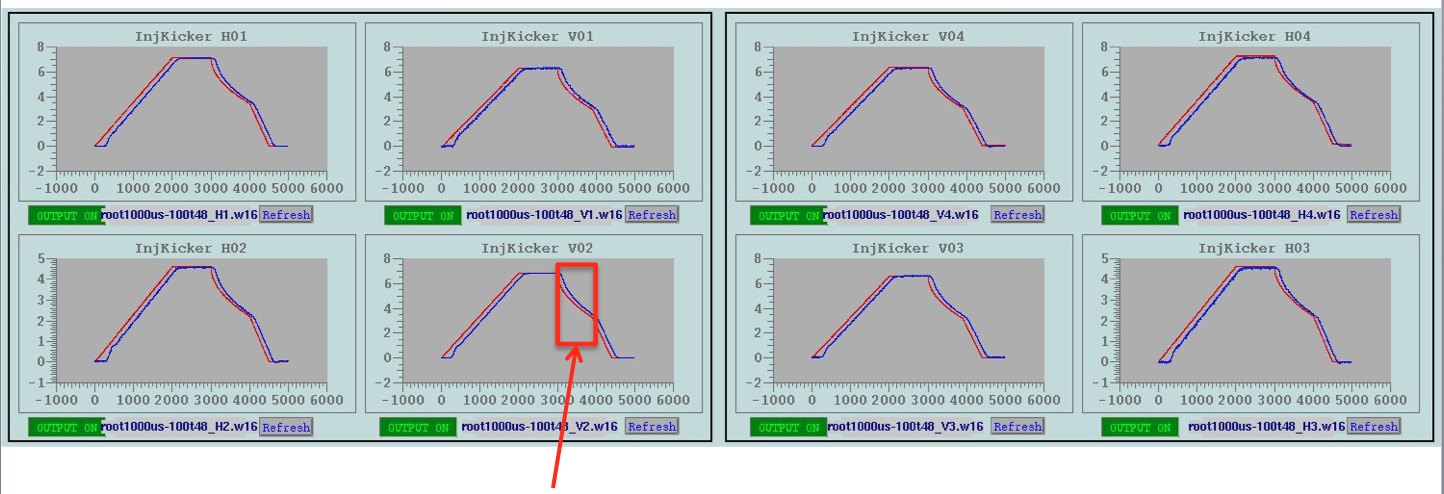}
\caption{\label{fig:inj_lockdown} (Colour)  A screen shot showing the injection kicker waveforms. The region marked by the rectangle and arrow shows the portion of the waveform that is monitored for one of the kickers (note that all eight are monitored).}
\end{center}
\end{figure}

There are four horizontal and four vertical injection kickers to paint the beam distribution in the ring.
An example set of waveforms is shown in  Fig.~\ref{fig:inj_lockdown}.
If for some reason the waveforms were accidentally set to paint a smaller beam, it would result in an excessively high beam density on the target.
Or, the kicker system may fail in a way that could also cause the beam size to be too small.
To protect against these possibilities an injection kicker waveform monitor system, based on commercially available oscilloscopes, watches the readback waveforms and automatically trips off the beam if the waveforms stray outside pre-determined envelopes.

Some interlocks are always in effect, not just when the beam power is greater than 100~kW.
Examples include the beam loss monitors (direct MPS interlock), the proton beam window halo thermocouples (shown in Fig.~\ref{fig:window}, and interlocked through a PLC connected to the MPS), and the target protection system, which monitors parameters such as mercury flow, water cooling, etc (PLC interlock).

\begin{figure}[!htb]
\begin{center}
\includegraphics [width=0.9\textwidth]{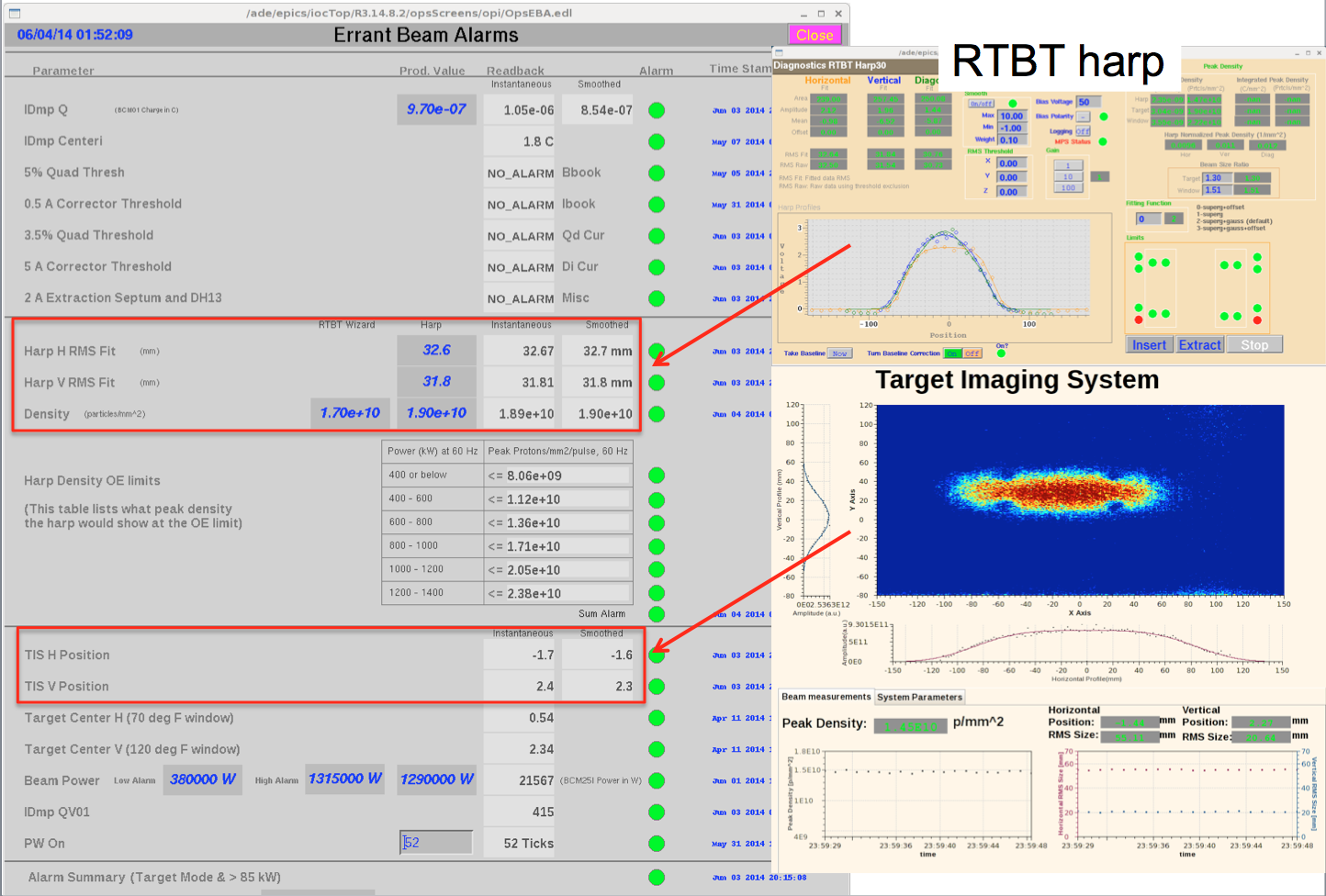}
\caption{\label{fig:alarm_summary} A screen shot of the alarm summary screen and example screen shots of the beam profile measured by the harp and the beam distribution measured by the TIS. The arrows show how the measured beam parameters are monitored by the alarm system.}
\end{center}
\end{figure}

The control system computers monitor the following parameters and alert the operators if they stray outside of pre-set bounds:
\begin{itemize}
\item beam position on target calculated by the target imaging system;
\item the beam density at the harp 9.52 m upstream of the target. The beam density at the target should be proportional to the beam density at the harp;
\item the beam size at the harp 9.52 m upstream of the target. The beam size at the target should be proportional to the beam size at the harp;
\item beam power;
\item beam centring estimated from the proton beam window halo thermocouples (top--bottom, left--right);
\item some magnet currents.
\end{itemize}
The last three items on the list are also interlocked by other methods as discussed above, but in this case the thresholds are more loosely set to give the operators a chance to correct the problem before it becomes severe enough to trip off the beam.
The alarm summary screen is shown in Fig.~\ref{fig:alarm_summary}.

%\begin{figure}[!htb]
%\begin{center}
%\includegraphics [width=0.8\textwidth]{fig_halo_tc.png}
%\caption{\label{fig:halo_tc} Photograph of the thermocouple beam halo monitors mounted to the proton beam window assembly. The ellipse shows the approximate beam size and location. Image reproduced from Ref. \cite{ref:McManamy2009b}.}
%\end{center}
%\end{figure}

%\FloatBarrier  %Uses the placeins package, prevents figures from previous sections to overflow into the following section.
\section{Summary}

In summary, high-power targetry is an active field that is continuously advancing.
The highest power targets require tight control over the beam position, density, and distribution, and even then the target lifetimes can be short (e.g. 6 months at SNS).
Machine protection systems must monitor these beam parameters and quickly activate interlocks if they stray outside of limits.
Protection can include monitoring equipment set points, equipment readback parameters, beam parameters, locking down certain controls, beam line design, and collimator systems.

\FloatBarrier  %Uses the placeins package, prevents figures from previous sections to overflow into the following section.
\section*{Acknowledgements}
ORNL is managed by UT-Battelle, LLC, under contract DE-AC05-00OR22725 for the U.S. Department of Energy.

Notice: This manuscript has been authored by UT-Battelle, LLC, under Contract No. DE-AC05-00OR22725 with the U.S. Department of Energy. The United States Government retains, and the publisher, by accepting the article for publication, acknowledges that the United States Government retains a non-exclusive, paid-up, irrevocable, world-wide licence to publish or reproduce the published form of this manuscript, or allow others to do so, for United States Government purposes.

%\bibliography{foils.bib}

\begin{thebibliography}{99}

\bibitem{ref:Haines2009} J. Haines, U.S. Particle Accelerator School, High Power Beam Targets Class, Vanderbilt University, Nashville, Tennessee, 20 January 2009.

\bibitem{ref:Reimer2014} B. Reimer, "Spallation Source Facilities," 5th High Power Targetry Workshop, Fermilab, 20--23 May 2014. https://indico.fnal.gov/conferenceDisplay.py?confId=7870. 

\bibitem{ref:Haynes2013} D. Haynes, "Introduction and Overview of the ISIS Target Systems," ISIS and SNS Bilateral Workshop, Oak Ridge, Tennessee, 25--26 June 2013.

\bibitem{ref:McManamy2009} T. McManamy, U.S. Particle Accelerator School, High Power Beam Targets Class, Vanderbilt University, Nashville, Tennessee, 20 January 2009. % ISIS TGT1 image, FNAL P-bar target image, side image of SNS PBW,

\bibitem{ref:Werbeck2003} R. Werbeck, "LANSCE Short-Pulse Target Operation," Workshop on High-power Targetry for Future Accelerators, Long Island, New York, 8--12 September 2003.  % LANSCE tgt image

\bibitem{ref:Wagner2011} W. Wagner et al., "The SINQ solid spallation target ?  operational experience and recent improvements ," High Power Targetry Workshop, Malmo, Sweden, 2--6 May 2011. Also http://www.hep.princeton.edu/mumu/target/Wagner/Wagner\_050311.pdf. % SINQ tgt.

\bibitem{ref:Mokhov2014} N. Mokhov, "Beam-Material Interactions," Joint US--CERN Accelerator School on Machine Protection, Newport Beach, CA, USA, 5--14 November 2014. % FNAL p-bar tgt mage, not quite the exact one I used but still a good ref.

\bibitem{ref:Hylen2014} J. Hylen, "Survey of Target Facility Landscape: Neutrino Beam Facilities," 5th High Power Targetry Workshop, Fermilab, 20--23 May 2014.  % FNAL NUMI tgt and J-PARC T2K tgt.

\bibitem{ref:Bricault2014} P. Bricault,  "Radioactive Ion Beam Facilities ? High Power Target, Current status and
Future Directions," 5th High Power Targetry Workshop, Fermilab, 20--23 May 2014. Also https://indico.fnal.gov/getFile.py/access?contribId=91\&sessionId=0\&resId=0\&materialId=slides\&\\confId=7870. % Both ISOL tgts.

\bibitem{ref:Hurh2007} P. Hurh, Third High Power Target Workshop, Bad Zurach, Switzerland, September 2007.

\bibitem{ref:McManamy2009b} T. McManamy, "Operational Experience with the SNS Target Systems and Upgrade Plans," Workshop on Applications of High Intensity Proton Accelerators, Fermi National Accelerator Laboratory, Batavia, IL, USA, 19--21 October 2009. % Photo of PBW

\bibitem{ref:Meigo2014a} S. Meigo, "Development of beam flattening system using non-linear beam optics at J-PARC," Eleventh International Topical Meeting on Nuclear Applications of Accelerators, Bruges, Belgium, 5--8 August 2014.

\bibitem{ref:Meigo2014b} S. Meigo et al., Beam flattening system based on non-linear optics for high power spallation neutron target at J-PARC, Proc. IPAC2014, Dresden, Germany, 15--20 June 2014.

\bibitem{ref:Thomsen2013} H.D. Thomsen, A.I.S. Holm, and S.P. M{\o}ller, "A linear beam raster system for the European Spallation Source," Proc. IPAC2013, Shanghai, China, May 2013, p. 70.

\bibitem{ref:jparcaccident} http://j-parc.jp/en/topics/20130812Accident\_Report.html; http://j-parc.jp/en/topics/HDAccident\\20131217.pdf

\bibitem{ref:ISISTGT2} L. Jones, PASI Working Group Report by D.M. Jenkins, April 2013; http://pasi.org.uk/images/2/2c/DavidJenkins\_PASI\_meeting\_28\_Feb\_13.pdf.

\bibitem{ref:Thomsen2007b} K. Thomsen and P.A. Schmelzbach, "A dedicated beam interrupt system for the safe operation of the Megapie Liquid Metal Target," Utilisation and Reliability of High Power Proton Accelerators (HPPA5), Workshop Proc., Mol, Belgium, 6--9 May 2007.

\bibitem{ref:Thomsen2007} K. Thomsen, "VIMOS, near-target beam diagnostics for MEGAPIE," {\em NIM A} \textbf{575}(3) (2007) 347. http://dx.doi.org/10.1016/j.nima.2007.03.011

\bibitem{ref:Sommer2003} W. F. Sommer et al., "Failure analysis of a radio-activated accelerator component," {\em Practical Failure Anal.} \textbf{3}(1) (2003) 71. http://dx.doi.org/10.1007/BF02717412

\bibitem{ref:Holmes2005} J. Holmes, "Quadrupole Strength Limits in the RTBT," SNS Tech. Note 162, 2005.

\bibitem{Catalan2001} N. Catalan-Lasheras and D. Raparia, "The Collimation System of the SNS Transfer Lines," Proc. of the 2001 Particle Accelerator Conference, Chicago, p. 3263. http://accelconf.web.cern.ch/AccelConf/p01/INDEX.HTM

\bibitem{Kinsho2014} M. Kinsho, Private communication, 2014.

\bibitem{Childress2008} S. Childress, "NUMI Proton Beam Diagnostics and Control: Achieving 2 Megawatt Capability," Proc. HB2008 Workshop, Nashville, TN, USA, 25--29 August 2008, p. 475.

\bibitem{Adams2014} D. Adams, Private communication, 2014.

\bibitem{ref:Plum2008} M. Plum, "SNS Injection and Extraction Systems - Issues And Solutions," Proc. HB2008 Workshop, Nashville, TN, USA, 25--29 August 2008, p. 268.

\end{thebibliography}

\end{document}